\documentclass{revtex4}

\usepackage{graphicx}
\usepackage{amsmath}
\usepackage{amsfonts}

\begin{document}

\title{New views on classical and quantum Brans-Dicke theory}

\author{J\'ulio C. Fabris$^{(a,b)}$,  Bertrand Chauvineau$^{(c)}$, Davi C. Rodrigues$^{(a)}$, Carla R. Almeida$^{(a)}$ and  Oliver F. Piattella$^{(a)}$}
\affiliation{$^{(a)}$Departamento de F\'{\i}sica - Universidade Federal do Esp\'{\i}rito Santo - Brasil}
\affiliation{$^{(b)}$National Research Nuclear University “MEPhI,” Kashirskoe sh. 31, Moscow 115409, Russia}
\affiliation{$^{(c)}$ Laboratoire Lagrange (UMR 7293), Universit\'e C\^ote d'Azur, CNRS,
Observatoire de la C\^ote d’Azur, France}

\begin{abstract}
The Brans-Dicke action is one of the most natural extensions of the Einstein-Hilbert action. 
It is based on the introduction of a fundamental scalar field that effectively incorporates a dynamics to the gravitational coupling $G$. In spite of the diverse motivations and the rich phenomenology that comes from its solutions, Solar System tests impose strong constraints on the Brans-Dicke theory, rendering it indistinguishable from General Relativity. 
In the present text, new perspectives for the Brans-Dicke theory are presented, based on the possibility that the scalar field presented in the BD theory can be external, as well as on the applications to black hole physics and the primordial universe.
\end{abstract}

\maketitle

\section{Introduction}

The present status of the General Relativity theory (GR) can be viewed under many different perspectives. From one side, GR is a very beautiful theory of gravitation, both mathematically and conceptually, which passes all local tests and experiments with flying colours. As an example, the Parametrised Post-Newtonian (PPN) parameters agree with the predictions of GR theory with a precision of about $10^{-5}$ \cite{will}. Due to the fact that the GR predictions are very small corrections of the usual Newtonian ones, these results may be seen as a strong evidence that GR could be considered as the correct classical theory of gravitation.

On the other side, however, GR suffers from many difficulties. First, it is plagued with singularity problems, either in the beginning of the history of the universe and in the final stage of the life of some massive stars. Moreover, in order to explain the observed dynamics and structure of the universe, the energy-matter sector must be constituted by $95\%$ of unknown components: dark matter (about $30\%$), necessary to explain the dynamics of virialized systems like galaxies and clusters of galaxies, as well as to allow for the structure formation process to take place; dark energy (about $70\%$), necessary to explain the present phase of accelerated expansion, as well as the features of the spectrum of the cosmic microwave background radiation (CMB), and to solve the age of the universe problem. To these issues we must add the difficulties in quantising GR, which constitute a serious limitation of the theory.

Many alternatives to GR have been proposed, and this is very important because the existence of competitive models may allow to better test the consistency of GR, which is considered as our standard gravitational model. Recently, there was a revival of such research for a more complete theory of gravitation, through $f(R)$ theories \cite{f(r)}, multidimensional theories \cite{duff}, theories with scalar fields with some specific properties, like the Horndesky \cite{horn} and Galileon theories \cite{galileon} (just to quote the most popular and promising recent proposals). However, the possible extensions/modifications of GR are a very old subject, even older than GR itself! Indeed, we may quote as the first tentative of describing gravity through a single scalar field, N\"ordstom's paper of 1912 \cite{st}.

One of the oldest proposal as an alternative to GR as a gravitational theory is the scalar-tensor theory, based on the non-minimal coupling of a scalar field to the gravitational sector. Somehow, it was initially based on the possibility that the gravitational coupling is not a constant, but may be a function of spacetime. This idea was initially evoked by Dirac \cite{dirac}, in the thirties, inspired by the coincidence between some representative numbers in physics, like the ratio between the gravitational and electric force between the electron and the proton in the hydrogen atom, and the characteristic time of the sub-atomic phenomena and the age of the universe, both giving numbers of about $10^{40}$. If this is not a coincidence, at least one of the fundamental constant of nature ($c$, $\hbar$ or $G$) may change with time, since the age of the universe is, of course, an increasing quantity.

The idea that the gravitational coupling may vary with time has been inserted into a consistent relativistic theory of gravity in the Brans-Dicke theory, at the beginning of the sixties \cite{bd}. There, the varying gravitational coupling is described through a long range scalar field $\phi$ non-minimally coupled to the usual Einstein-Hilbert Lagrangian, which is the base of GR. The structure of the theory implies the presence of a new (dimensionless) coupling constant $\omega$ such that when $\omega \rightarrow \infty$ the GR theory, with a constant gravitational coupling, is recovered. This theory tries also to insert Mach's principle in a geometric gravitational theory.

The introduction of a non-minimal coupled scalar field, whose inverse is connected with the value of $G$, leads to many new possibilities. However, the use of PPN estimations implies, in the absence of a potential term, that $\omega > 40.000$ \cite{will}. This makes Brans-Dicke theory almost indistinguishable from GR, and for this reason the interest in that theory has considerably decreased. Remark that this very strict constraint is established by local gravitational test, whereas cosmology implies less restrictive bounds on $\omega$ \cite{cosmo}. However, from the theoretical point of view, the BD theory remained an interesting proposal: when $\omega = - 1$, it coincides with the low energy effective action coming from string theory, in its simplest version; its structure may be mapped into multidimensional theories, which is a current domain of research.

Is it possible to revive the traditional BD theory by looking at it in a different way? If yes, how to reconcile with the very constraining results coming from local experiments/observations (which give rise the estimations of the PPN parameters, for example)? This is the subject of the present paper. We will focus on these questions first by showing that the BD theory exhibits some features not yet completely explored, and then presenting an approach which can, potentially, reconcile the interesting new features of the BD theory with the PPN results. This approach is based on the possibility that the scalar field presented in the BD theory can be external (or, within another terminology, non-dynamical \cite{wei}), meaning that it is not subject to a variational principle, its dynamics being dictated by consistency conditions of the field equations represented by Bianchi's identities. Finally, we will sketch some properties of the quantisation of the Brans-Dicke theory in the mini-superspace, which seems to favour negative values of $\omega$.

Professor Spindel has worked intensively on fundamental aspects of physics, in particular gravitational physics, and we think that such {\it variations} on the Brans-Dicke theme is a good way to honour him in the occasion of his $65^{th}$ birthday.
 
\section{Scalar fields}

The introduction of scalar fields is the simplest way to go beyond General Relativity, since it adds a spin $0$ degree of freedom to the theory. We can think of many different possibilities of introducing a scalar field in a gravitational context. We will revise very briefly some of them.

The first one makes use of the minimal coupling between the scalar field and gravitation, and it is defined by the Lagrangian:
\begin{eqnarray}
{\cal L} = \sqrt{-g}\biggr\{R - \epsilon\phi_{;\rho}\phi^{;\rho} + 2V(\phi)\biggl\}\;.  
\end{eqnarray}
In this Lagrangian, $V(\phi)$ is the potential self-interacting term. The corresponding energy-momentum tensor satisfies all the energy conditions if $\epsilon = 1$ (ordinary field) but violates the strong and null energy conditions if $\epsilon = - 1$ (phantom fields).

We may have a much less trivial configuration using the non-minimal coupling:
\begin{eqnarray}
{\cal L} = \sqrt{-g}\biggr\{f(\phi)R - \omega(\phi)\frac{\phi_{;\rho}\phi^{;\rho}}{\phi} + 2V(\phi)\biggl\}\;, 
\end{eqnarray}
where $f(\phi)$ and $\omega(\phi)$ are, in general, functions of the scalar field itself. The case
\begin{eqnarray}
f(\phi) = \phi, \quad \omega(\phi) =\mbox{constant}, \quad V(\phi) = 0\;, 
\end{eqnarray}
defines the traditional Brans-Dicke theory, which is considered as the prototype of the scalar-tensor theories:
\begin{eqnarray}
\label{bd}
{\cal L} = \sqrt{-g}\biggr\{\phi R - \omega\frac{\phi_{;\rho}\phi^{;\rho}}{\phi} )\biggl\} + {\cal L}_m(g_{\mu\nu},\Psi)\;,
\end{eqnarray}
where $\Psi$ represents the matter degrees of freedom, which couple minimally to the metric.

Another possibility of including scalar fields is a non-canonical kinetic term, leading to the $K$-essence models:
\begin{eqnarray}
{\cal L} = \sqrt{-g}\biggr\{R - \omega(\phi)f(X)+ 2V(\phi)\biggl\}\;, 
\end{eqnarray}
where
\begin{eqnarray}
X = \phi_{;\rho}\phi^{;\rho}\;,  
\end{eqnarray}
and $f$ is a given function of $X$.

Other possibilities of constructing a scalar-tensor theory include, for example, the Horndeski \cite{horn} theory, which is the most general scalar-tensor Lagrangian leading to second order differential equations. Another interesting proposal, which is related to the Horndeski class of theory, is the Galileon theory \cite{galileon}, where a scalar field exhibits some special translational symmetries. One example of such generalisation is the theory defined by:
\begin{eqnarray}
{\cal L} = \sqrt{-g}\biggr\{R - G^{\mu\nu}\phi_{;\mu}\phi_{;\nu}+ 2V(\phi)\biggl\}\;, 
\end{eqnarray}
where $G_{\mu\nu}$ is the Einstein tensor.

In what follows we limit ourselves to the traditional Brans-Dicke theory, represented by the action (\ref{bd}). The corresponding field equations are:
\begin{eqnarray}
R_{\mu\nu} - \frac{1}{2}g_{\mu\nu}R &=& \frac{8\pi}{\phi} T_{\mu\nu} + \frac{\omega}{\phi^2}(\phi_{;\mu}\phi_{;\nu} 
- \frac{1}{2}g_{\mu\nu}\phi^{;\rho}\phi_{;\rho}) + \frac{1}{\phi}(\phi_{;\mu;\nu} - g_{\mu\nu}\Box\phi)\;,\\
\Box\phi &=& \frac{8\pi}{3 + 2\omega}T\;,\\
{T^{\mu\nu}}_{;\mu}  &=& 0\;.
\end{eqnarray}
Remark that the usual conservation law for the matter component is preserved in the traditional Brans-Dicke theory. This is a property related to the preservation of the general diffeomorphism invariance in this theory.

\section{Static, spherically symmetric solutions}

In what follows we will consider the solutions describing a vacuum static spherically symmetric configuration. Such symmetry can represent a static star or an eternal black hole. For the cases where the rotation of these objects is small, this symmetry is usually a very good approximation \cite{frolov, chandra}. In order to inspecting better the situation, we perform a conformal transformation in the action (\ref{bd}):
\begin{eqnarray}
g_{\mu\nu} = \phi^{-1}\tilde g_{\mu\nu}\;, 
\end{eqnarray}
and write:
\begin{eqnarray}
\frac{d\phi}{d\sigma} = \biggr|\frac{3 + 2\omega}{\phi^2}\biggl|^\frac{1}{2}\;, 
\end{eqnarray}
With these redefinitions, we end up with the following Lagrangian:
\begin{eqnarray}
{\cal L} = \sqrt{- g}\biggr\{R - \epsilon\sigma_{;\rho}\sigma^{;\rho}\biggl\}, \quad \epsilon = \pm 1\;. 
\end{eqnarray}
Hence, the conformal transformation has allowed us to pass from the Jordan's frame (for which the scalar field couples non-minimally to the Einstein-Hilbert term) to the Einstein's frame (for which the scalar field is minimally coupled to the Ricci scalar). In principle, this operation is just a redefinition of the variables of the theory (including the metric). However, the interpretation of some aspects of the theory becomes easier in the Einstein's frame. For example, the scalar field has a clear kinetic energy contribution in this frame; the same does not happens in the Jordan's frame due to the non-minimal coupling. Remark that the conformal transformation affects also the matter sector and leads to a non-minimal coupling between the matter degrees of freedom and the scalar field.

One important feature is that the parameter $\epsilon$ has a positive value if $\omega > - \frac{3}{2}$, while it is negative if $\omega < - \frac{3}{2}$. In the first case, the scalar field has the usual expression for the kinetic term, and it is called a {\it ordinary scalar field}, while in the second case the sign of the kinetic term is negative and the scalar field is usually called a {\it phantom scalar field} \cite{cald}. In fact, if the kinetic energy appears with the {\it wrong} sign, the null energy condition is violated. This property of the phantom scalar field may lead to instabilities, at least at the quantum level. However, interesting new features appear when we quantise the Brans-Dicke theory, at least in the mini-superspace. In fact, at some simplified quantum formulation, in the mini-superspace, the phantom scalar field has some nice properties rendering, for example, the effective Hamiltonian self-adjoint; the same does not necessarily happen for the ordinary scalar field \cite{carla}, as it will be discussed later. 

A class of static, spherically symmetric solutions for the Brans-Dicke theory in the vacuum is given by \cite{kirill}:
\begin{eqnarray} 
ds^2 = P^{-\xi}\biggr\{P^adt^2 - P^{-a}du^2 - P^{1 - a}u^2d\Omega^2\biggl\}\;, 
\end{eqnarray}
with 
\begin{eqnarray} 
P(u) = 1 - 2\frac{k}{u} \quad &,& \phi = P^\xi\;,  \\
\frac{b}{k} = a \quad &,& \quad a^2 = 1 - (3 + 2\omega)\xi^2\;, \\
2k^2\mbox{sign} k &=& 2b^2 + \epsilon C^2\;. 
\end{eqnarray}
Black hole solutions exist only when $3 + 2\omega < 0$, corresponding to the phantom configuration in the Einstein's frame. In fact, if $3 + 2\omega > 0$, $a < 1$ and the possible horizon at $u = 2k$ is a center. However, in the phantom case, $a > 1$, and $u = 2k$ can represent a horizon, even if with very peculiar properties.

At this point it is important to stress that if the scalar field has not a phantom nature (in the Einstein's frame), no black hole solution can be obtained, but only naked singularity solutions \cite{fisher, ellis, bronnikov}. This is true for a minimally coupled scalar field \cite{nelson} and for the traditional Brans-Dicke theory \cite{kirill}. However, such restriction can be relaxed in other configurations as the string-inspired effective action in the low energy limit \cite{clement1, clement2}.

As a matter of fact, we can wonder if the solutions described above are indeed black holes, due to their unusual properties. Among the latter, we can quote the following:
\begin{enumerate}
\item Their surface gravity is zero. It means that their Hawking temperature (if it is possible to define it, see Ref. \cite{brasil-bis}) is zero. For this reason, these black holes has been called {\it Cold black holes}.

\item The surface of the event horizon is infinite. This implies an infinite entropy by the area law. But, if the temperature is zero, the entropy cannot be in principle infinite. Hence, the computation of the entropy in this case may not follow the classical lines devised for the black holes emerging from GR.
 
\item The tidal forces are infinite over the event horizon hypersurface. But, point particles can cross the latter.

\item The solutions are asymptotically flat. Hence, static observers at the infinity can be defined.
\end{enumerate}
There is another important feature: the solutions contain a scalar charge, hence these solutions contain hair. This fact points out the limitations of the so-called {\it no-hair theorem}, which seems to be restricted to very specific situations.

In fact, the general solution for these scalar black holes reveals two types of asymptotically flat black hole, besides the Schwarzschild one \cite{kirill}: $i)$ type B1 black holes, where the horizon can be crossed in a finite proper time by an in falling particle; $ii)$ type B2 black holes, where the horizon is reached in a infinite proper time for an infalling particle. In both cases, there is always a throat at a finite distance of the horizon.

The stability of these structures is more controversial. Some methods of computing the perturbations lead to the conclusion of stability, but other methods lead to the opposite conclusion \cite{kirill, mexico, zhidenko}.

\section{Cosmology}

The Brans-Dicke equations admit very simple solutions for an isotropic and homogeneous universe, described, for example, by the flat Friedmann-Lema\^{\i}tre-Robertson-Walker metric:
\begin{eqnarray}
ds^2 = dt^2 - a(t)^2(dx^2 + dy^2 + dz^2)\;.
\end{eqnarray}
The equations read:
\begin{eqnarray}
3H^2 &=& \frac{8\pi}{\phi}\rho + \frac{\omega}{2}\biggr(\frac{\dot\phi}{\phi}\biggl)^2 - 3H\frac{\dot\phi}{\phi}\;, \\
2\dot H + 3H^2 &=& - \frac{8\pi}{\phi}p - \frac{\omega}{2}\biggr(\frac{\dot\phi}{\phi}\biggl)^2 - \frac{\ddot\phi}{\phi}
- 2 H\frac{\dot\phi}{\phi}\;,\\
\ddot\phi + 3H\dot\phi &=& \frac{8\pi}{3 + 2\omega}(\rho - 3p)\;,\\
\dot\rho + 3H(\rho + p) &=& 0\;.
\end{eqnarray}
These equations admit quite simple solutions for some specific equations of state. 
\begin{enumerate}
\item For a radiative universe, with $T^\rho_\rho = 0$, a simple solution implies $\dot\phi = 0$. This gives the usual radiative solution of GR theory, with $a \propto t^\frac{1}{2}$. This is quite attractive, since the fact the radiative phase of the standard cosmological model can be reproduced, implies that a BD cosmological scenario can preserve the results from nucleosynthesis and those connected with the relic radiation that permeates the universe today.

\item For a matter dominated universe, with $p = 0$, one obtains the power law solutions:
\begin{eqnarray}
\label{s1}
a &=& a_0 t^\frac{2 + 2\omega}{4 + 3\omega}\;,\\
\label{s2}
\phi &=& \phi_0 t^\frac{2}{4+ 3\omega}\;.
\end{eqnarray}

\item If the universe is dominated by the vacuum energy density, with $p = - \rho$, implying that $\rho = $ constant by using the conservation law, there is also a power law solution given by:
\begin{eqnarray}
a &=& a_0 t^{\omega + \frac{1}{2}}\;,\\
\phi &=& t^2\;.
\end{eqnarray}
This solution has been evoked in Ref. \cite{infla-bd}, leading to a revival of the old inflationary scenario \cite{guth}. However, the transition to the radiative phase, together with the necessary number of $e$-folds to solve the problems of the standard cosmological model, can be achieved only if $\omega \sim 20$, in disagreement with the PPN results.
\end{enumerate}

One interesting aspect of the solutions during the matter-dominated phase is that they admit scenarios for which the universe is accelerating if $- \frac{3}{2} < \omega < - \frac{4}{3}$. However, in this range for the parameter $\omega$ we have at least three problems: $i)$ it disagrees with PPN limits quoted in the Introduction; $ii)$ it is a eternal accelerating regime, and this implies that the structure formation process would not be possible in such a context; $iii)$ it implies a negative effective gravitational coupling. Hence, in principle, it seems not possible to implement a realistic scenario. 

We further study the issue of matter-dominated solutions in the BD cosmology in the next section.

\section{A general spatially flat solution in the Brans-Dicke theory}

In Ref. \cite{gurevich}, the authors have determined the general solution for a spatially flat, isotropic and homogeneous universe with a general linear equation of state, which includes the main equations of state of interest to cosmology: radiative phase, matter-dominated phase, stiff matter and cosmological constant phase. In what follows, we will concentrate ourselves on the general flat solutions for the pressure-less matter-dominated era. They read as follows:
\begin{enumerate}
\item $\omega > - \frac{3}{2}$ :
\begin{eqnarray}
\label{sa1}
a &=& a_0(t - t_+)^\frac{1 + \omega \pm \sqrt{1 + \frac{2}{3}\omega}}{4 + 3\omega}(t - t_-)^\frac{1 + \omega \mp \sqrt{1 + \frac{2}{3}\omega}}{4 + 3\omega}\;,\\
\label{sa2}
\phi &=& \phi_0(t - t_+)^\frac{1 \pm 3\sqrt{1 + \frac{2}{3}\omega}}{4 + 3\omega}(t - t_-)^\frac{1 \mp 3\sqrt{1 + \frac{2}{3}\omega}}{4 + 3\omega}\;,
\end{eqnarray}
\item $\omega < - \frac{3}{2}$ :
\begin{eqnarray}
\label{sb1}
a &=& a_0[(t + t_-)^2 + t_+^2]^\frac{1 + \omega}{4 + 3\omega}\times \exp\biggr\{\pm \frac{\sqrt{- \frac{2}{3}\omega - 1}}{4 + 3\omega}\arctan\biggr(\frac{t + t_-}{t_+}\biggl)\biggl\}\;,\\
\label{sb2}
\phi &=& \phi_0(t + t_-)^2 + t_+^2]^\frac{1 + \omega}{4 + 3\omega}\times \exp\biggr\{\mp 3\frac{\sqrt{- \frac{2}{3}\omega - 1}}{4 + 3\omega}\arctan\biggr(\frac{t + t_-}{t_+}\biggl)\biggl\}\;.
\end{eqnarray}
\end{enumerate}
In these expressions, $t_\pm$ are integration constant, and we assume that $t_+ > t_-$.

These solutions have some striking properties. Let us first consider the case $\omega > - \frac{3}{2}$ represented by the expressions (\ref{sa1}) and (\ref{sa2}). Since $t_+ > t_-$, the initial singularity occurs at $t = t_+$. Hence, when $t \rightarrow t_+$, the solution reads:
\begin{eqnarray}
\label{asymptote}
a \propto (t - t_+)^\frac{1 + \omega \pm \sqrt{1 + \frac{2}{3}\omega}}{4 + 3\omega}\;.
\end{eqnarray}
On the other hand, when $t \rightarrow \infty$, we can neglect $t_\pm$ and the solution takes the form given by(\ref{s1}). There are some interesting properties concerning these asymptotic behaviours, when $- \frac{3}{2} < \omega < - \frac{4}{3}$. In fact, in this case, initially the solution represents a decelerating universe, while it becomes an accelerating expansion in the late-times regime: the decelerating/accelerating transition is achieved naturally. This model has been tested in Ref. \cite{fabris} using type Ia supernovae data, with slightly better results than the usual $\Lambda$CDM model. However, the price to pay is quite big: in the given interval $- \frac{3}{2} < \omega < - \frac{4}{3}$ the gravitational coupling is negative. The most dangerous aspect, in our point of view, is that such gravitational repulsive phase may be unstable against small perturbations.

However, the problems pointed out above does not exhaust the richness of such solution. In fact, (\ref{s1}) implies three possibilities (we exclude for the moment the case $\omega < - \frac{3}{2}$): $i)$ if $- \frac{3}{2} < \omega < - \frac{4}{3}$, the solution exhibits an accelerated expansion, but with a negative gravitational coupling; $ii)$ if $- \frac{4}{3} < \omega < - 1$, the exponent is negative, and it is also an accelerated universe, but with positive gravitational coupling; $iii)$ if $- 1 < \omega$, the solution represents a decelerating expanding universe.

In the other asymptote, there are much more possibilities. Let us consider the behaviour dictated by the asymptotical expression (\ref{asymptote}). If the sign plus is chosen in the numerator, the universe is expanding and decelerating. However, if the sign minus is chosen, three possibilities are obtained: $i)$ if $- \frac{3}{2} < \omega < - \frac{4}{3}$ the expansion is accelerated; $ii)$ if $- \frac{4}{3} < \omega < 0$, the exponent is negative, and the second derivative of the scale factor is positive; $iii)$ if $\omega  > 0 $ the expansion is decelerating.

In figure 1, we plot the value of exponent of these general flat solutions in their asymptotical regions. In the asymptote where the scale factor behaves as (\ref{s1}), there is accelerated expansion in the region $ - \frac{4}{3} < \omega < - 1$, while in the other asymptote the behaviour depends on the sign in front of the square root in (\ref{asymptote}), as described earlier.

If $\omega < - \frac{3}{2}$, the main characteristic is the absence of singularities. In fact, in this case the solution behaves typically as displayed in figure 2. When $\omega < - \frac{3}{2}$ there is a phantom scalar field in the minimally coupled version of the Brans-Dicke theory. However, the gravitational coupling can be positive, at least for $\omega < - 2$. Hence, at classical level it is possible that this configuration can be stable.

\begin{figure}[!t]
\begin{center}
\begin{minipage}[t]{0.3\linewidth}
\includegraphics[width=\linewidth]{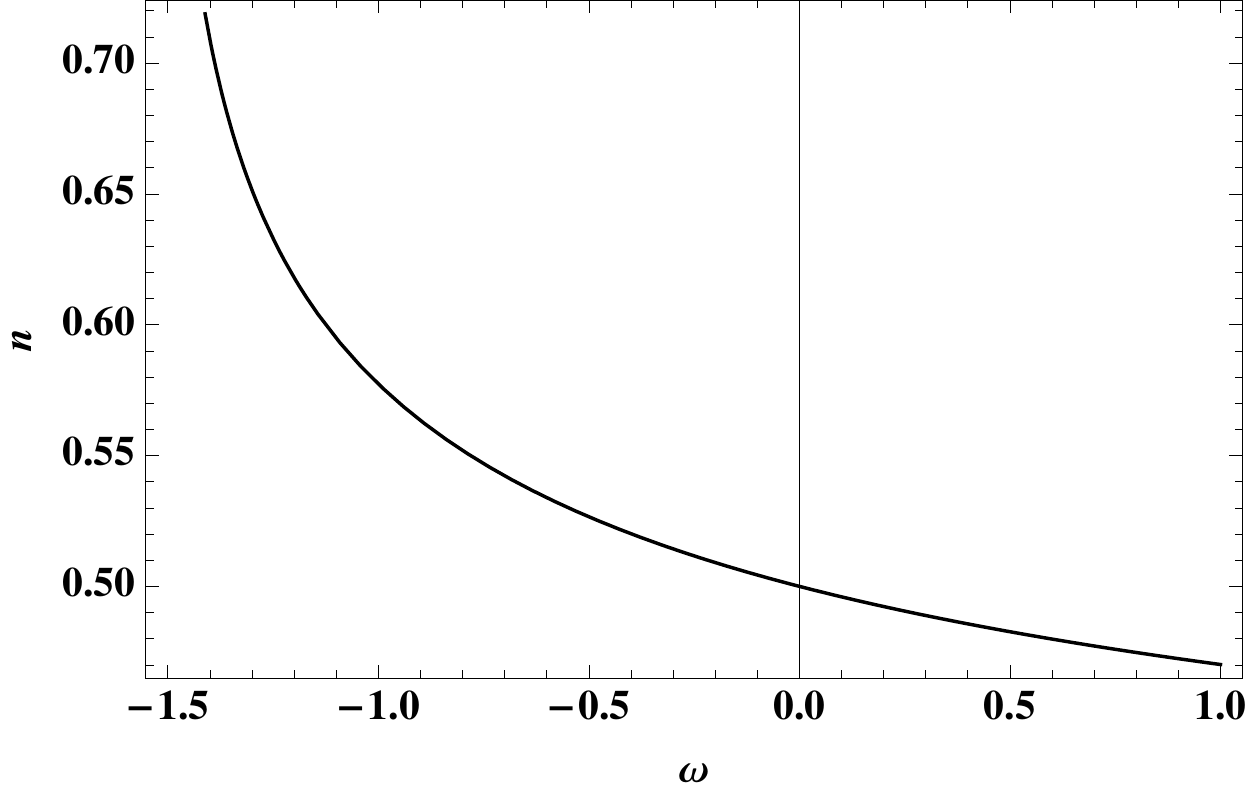}
\end{minipage} \hfill
\begin{minipage}[t]{0.3\linewidth}
\includegraphics[width=\linewidth]{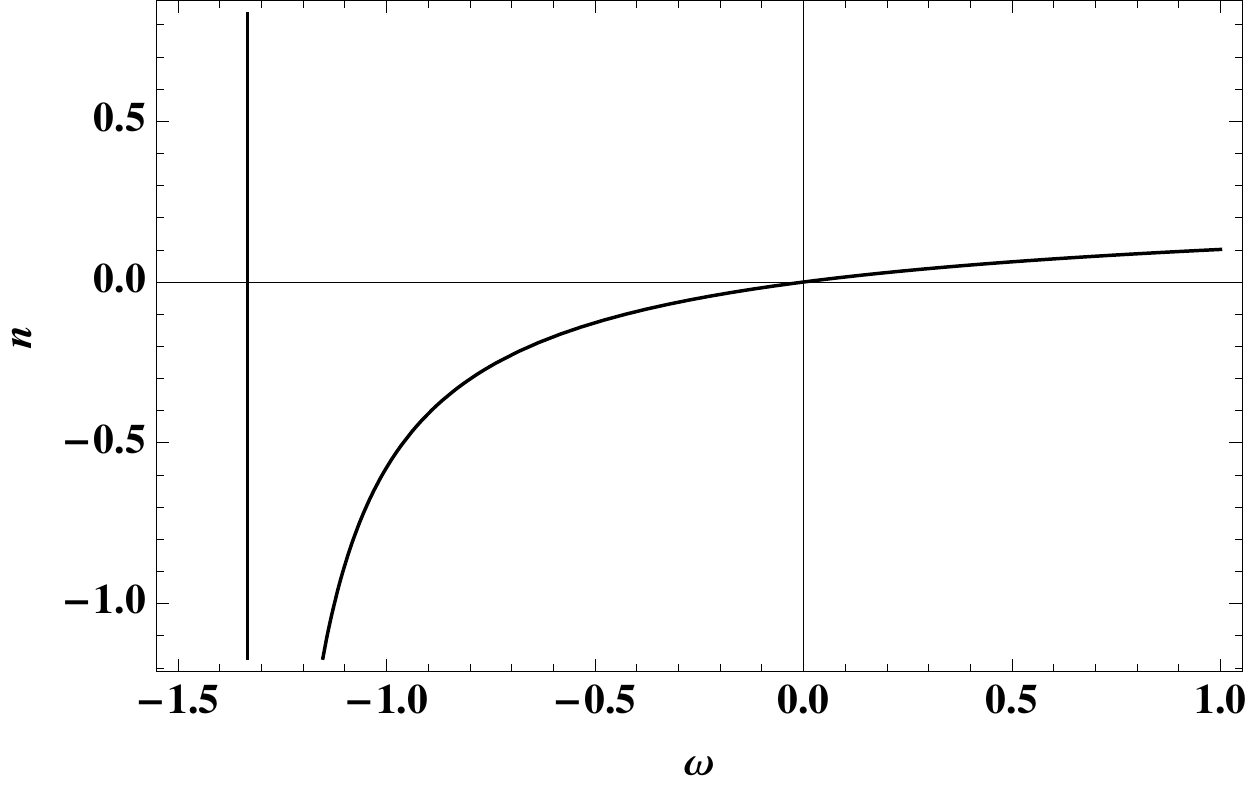}
\end{minipage} \hfill
\begin{minipage}[t]{0.3\linewidth}
\includegraphics[width=\linewidth]{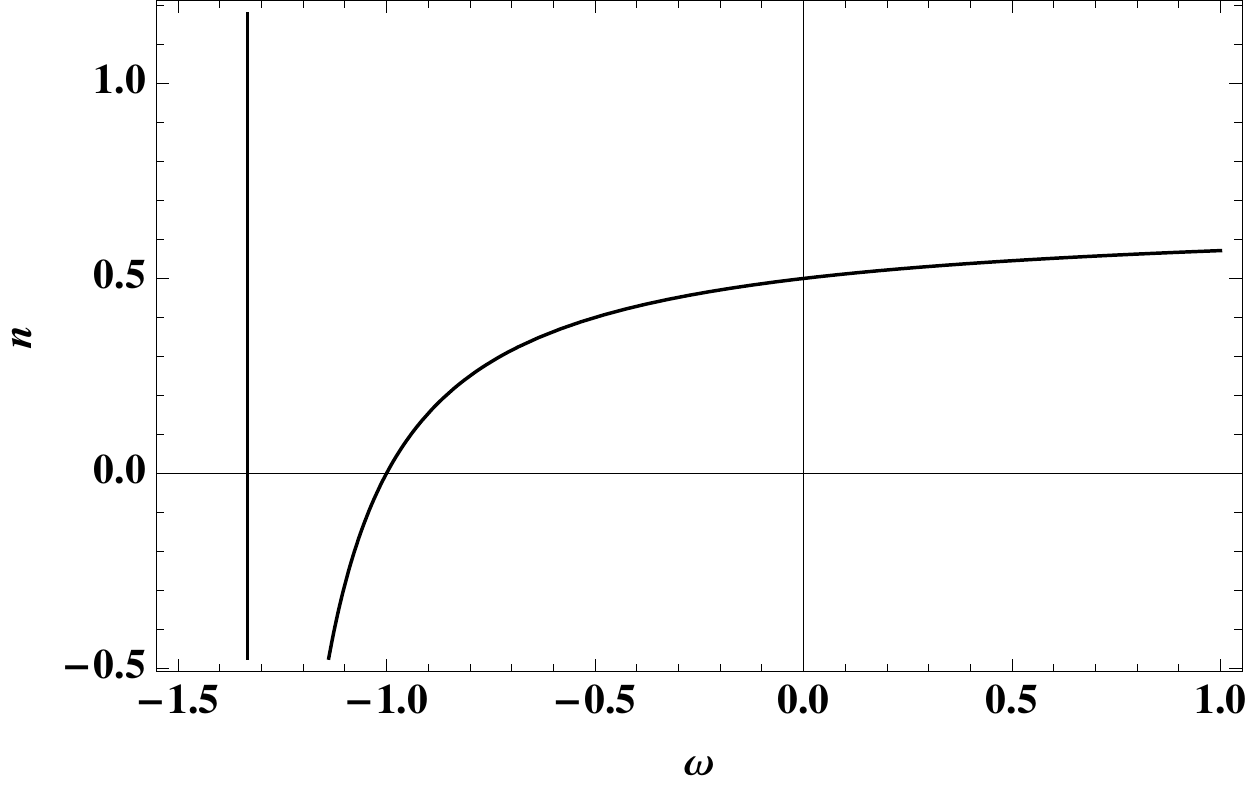}
\end{minipage} \hfill
\end{center}
\caption{The value of the exponent $n$ in the expression for the scale factor $a \propto t^n$, for the different asymptotical regions. In the left, when the term $n = (1 + \omega + \sqrt{1 + \frac{2}{3}\omega})/(4 + 3\omega)$ dominates. In the center when $n = (1 + \omega - \sqrt{1 + \frac{2}{3}\omega})/(4 + 3\omega)$ dominates. In the right the exponent of the other asymptote, $n = (2 + 2\omega)/(4 + 3\omega)$.}
\end{figure}

\begin{figure}[!t]
\begin{center}
\begin{minipage}[t]{0.5\linewidth}
\includegraphics[width=\linewidth]{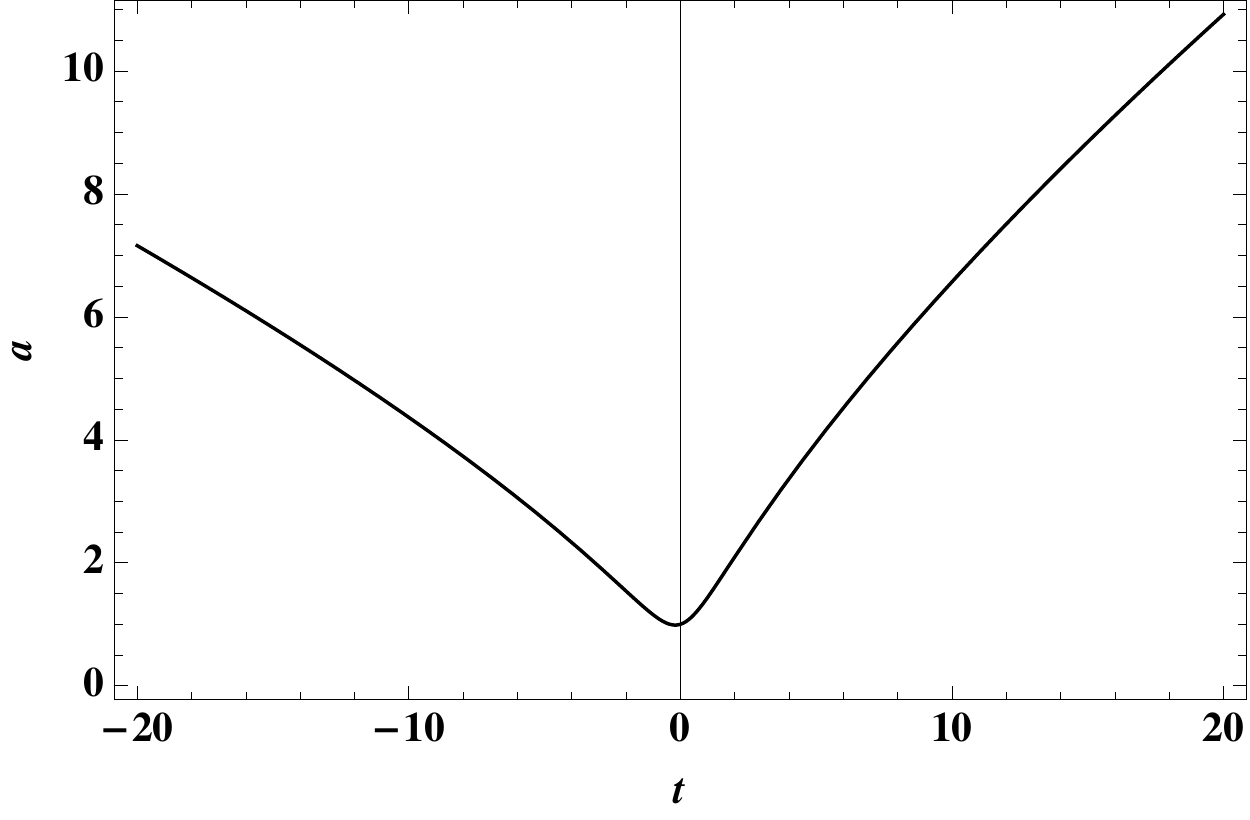}
\end{minipage} \hfill
\end{center}
\caption{Typical behaviour of the scale factor for $\omega < - 3/2$. In this case, $\omega = - 5$.}
\end{figure}

\section{External field}

The results presented in the previous section suffers from a major drawback: even if we can obtain very good results with the Brans-Dicke theory, the PPN analysis based on the local tests lead to a huge value for $\omega$ and this implies that the BD theory is very close to the GR theory. In this section, we discuss a mechanism that may turn the BD competitive even for small values of $\omega$. This mechanism is based on the concept of external field, that is defined in what follows.

We can define an {\it external field} as a field not subject to the variational principle. In the process of deriving the field equations from an action, the variation with respect to all the fields are considered, except for the external ones. In particular, this procedure is justified if the expression of the external field as a space-time function is known beforehand. An external field is also called sometimes as a {\it non-dynamical field} \cite{wei}. One of the relevance of the external field approach is the fact that it may be a procedure to include in a given system defined by a Lagrangian the influence of external factors. It is somehow connected with the principle of {\it open systems}. One simple exemple is the problem of a trajectory of a charge submitted by a constant electric field.

We present a specific example of how implementing the idea of external fields, based on reference \cite{chauvineau}. Let us consider the Brans-Dicke action, but with the scalar field being external. Moreover, we consider a cosmological term depending on this external scalar field.
\begin{eqnarray}
{\cal L} = \sqrt{-g}\biggr\{\phi R - \omega(\phi)\frac{\phi_{;\rho}\phi^{;\rho}}{\phi }- 2\phi\Lambda(\phi)\biggl\} + {\cal L}_m\;.
\end{eqnarray}
It has been also assumed that the Brans-Dicke coupling may depend on the scalar field. The field equations are now:
\begin{eqnarray}
R_{\mu\nu} - \frac{1}{2}g_{\mu\nu}R + g_{\mu\nu}\Lambda &=& \frac{8\pi}{\phi}T_{\mu\nu} + \frac{\omega}{\phi^2}(\phi_{;\mu}\phi_{;\nu} - \frac{1}{2}g_{\mu\nu}\phi_{;\rho}\phi^{;\rho})
+ \frac{1}{\phi}(\phi_{;\mu;\nu} - g_{\mu\nu}\Box\phi)\;,\\
{T^{\mu\nu}}_{;\mu} &=& 0\;.
\end{eqnarray}
The conservation of the energy-momentum tensor is still valid because of diffeomorphism invariance: the matter Lagrangian does not depend on the external field. There is not, in principle, any specific equation for the scalar field, since it is an external field not subject to the variational principle.

On the other hand, Bianchi's identities are still valid, leading to
\begin{eqnarray}
\biggr\{\Box\phi - \frac{1}{3 + 2\omega}\biggr[8\pi T + \omega'\phi_{;\rho}\phi^{;\rho} + 2(\Lambda - \phi\Lambda')\biggl]\biggl\}\phi_{;\mu} = 0\;,
\end{eqnarray}
where the primes mean derivative with respect to the scalar field $\phi$. The very important difference with respect to the usual Brans-Dicke theory is that now, $\phi = \mbox{constant}$ is a solution for any $\omega$ and even in presence of matter. This is not possible in the usual Brans-Dicke theory.

We may apply this construction to cosmology. Let us consider the flat FLRW metric:
\begin{eqnarray}
ds^2 = dt^2 - a(t)^2(dx^2 + dy^2 + dz^2)\;.
\end{eqnarray}
For simplicity, we ignore the cosmological term $\Lambda$ and we take $\omega$ as a constant. Moreover, let us suppose a pressureless matter content. The equations of motion are the following:
\begin{eqnarray}
3\biggr(\frac{\dot a}{a}\biggl)^2 &=& \frac{8\pi \rho}{\phi} + \frac{\omega}{2}\frac{\dot\phi^2}{\phi^2} - 3\frac{\dot a}{a}\frac{\dot\phi}{\phi}\;,\\
2\frac{\ddot a}{a} + \biggr(\frac{\dot a}{a}\biggl)^2 &=& - \frac{\omega}{2}\frac{\dot\phi^2}{\phi^2} - 2\frac{\dot a}{a}\frac{\dot\phi}{\phi} - \frac{\ddot\phi}{\phi}\;,\\
\bigg\{\ddot\phi + 3\frac{\dot a}{a}\dot\phi - \frac{8\pi\rho}{3 + 2\omega}\biggl\}\dot\phi &=& 0\;,\\
\dot\rho + 3\frac{\dot a}{a}\rho &=& 0\;.
\end{eqnarray}
The solutions for General Relativity (flat case), in the matter dominated phase are given by:
\begin{eqnarray} 
a(t) = A(t - t_0)^{2/3},\quad \phi &=& C\;.
\end{eqnarray}
The general flat Brans-Dicke solutions for $p = 0$ are:
\begin{eqnarray} 
a(t) &=& B(t - t_+)^{q_+}(t - t_-)^{q_-}\;,\\
\phi(t) &=& \phi_0(t - t_+)^{p_+}(t - t_-)^{p_-}\;,
\end{eqnarray}
with
\begin{eqnarray}
q_{\pm} &=& \frac{1 + \omega \pm \sqrt{1 + \frac{2}{3}\omega}}{4 + 3\omega}\;,\\
p_{\pm} &=& \frac{1 \pm 3\sqrt{1 + \frac{2}{3}\omega}}{4 + 3\omega}\;,
\end{eqnarray}
and $t_+ > t_-$. In order to have positive energy and positive gravitational coupling we must have $ \omega > - \frac{4}{3}$. These Brans-Dicke solutions were described in reference \cite{gurevich}.

The GR and BD solutions can not be matched in the traditional context: they are two different theories. The situation changes if $\phi$ is now an external field. Let us call such theory, {\it EST} for External Scalar Theory. In this case, the above described GR solution is also a solution for gravitation-external field system. Hence, patchwork solutions, constructed with regions where GR are valid together with other regions where BD solutions are valid, become possible.

Let us match the solutions in the external field approach. We fix $t_0 = 0$, and the transition time $t = t_m > t_+$. The solutions for the matter component read:
\begin{eqnarray}
6\pi \epsilon _{GR}\left( t\right)  &=&\frac{C}{t^{2}}\;,
\label{dust densities} \\[.1in]
4\pi \epsilon _{BD}\left( t\right)  &=&\frac{3+2\omega }{4+3\omega }\frac{%
\Phi _{0}}{\left( t-t_{+}\right) ^{3q_{+}}\left( t-t_{-}\right) ^{3q_{-}}}\;. 
\end{eqnarray}%
The subscripts $GR$ and $BD$ stand for the different phases, the GR phase and the BD phase, respectively. The continuities of the scalar and its (logarithmic) derivative, of the scale factor and its (logarithmic) derivative, give respectively
\begin{equation}
\Phi _{0}\left( t_{m}-t_{+}\right) ^{p_{+}}\left( t_{m}-t_{-}\right)
^{p_{-}}=C\;,  \label{scalar cont}
\end{equation}
\begin{equation}
\frac{p_{+}}{t_{m}-t_{+}}+\frac{p_{-}}{t_{m}-t_{-}}=0\;,
\label{scalar-der cont}
\end{equation}
\begin{equation}
B\left( t_{m}-t_{+}\right) ^{q_{+}}\left( t_{m}-t_{-}\right)
^{q_{-}}=At_{m}^{2/3}\;,  \label{scalefact cont}
\end{equation}
\begin{equation}
\frac{q_{+}}{t_{m}-t_{+}}+\frac{q_{-}}{t_{m}-t_{-}}=\frac{2}{3t_{m}}\;.
\label{scalefact der cont}
\end{equation}
The final expressions are:
\begin{eqnarray}
& &t_{\pm }=\frac{1}{1\pm 3\sqrt{1+\frac{2}{3}\omega }}t_{m}\;,\\
& &B =A\left( 3\sqrt{1+\frac{2}{3}\omega }+1\right) ^{q_{+}}\left( 3\sqrt{1+%
\frac{2}{3}\omega }-1\right) ^{q_{-}}\nonumber\\
&\times&\left( 9+6\omega \right) ^{-\frac{%
1+\omega }{4+3\omega }}t_{m}^{\frac{2}{3\left( 4+3\omega \right) }}\;,
\label{BD param Bphi} \\
& &\Phi _{0} = C\left( 3\sqrt{1+\frac{2}{3}\omega }+1\right) ^{p_{+}}\left( 3%
\sqrt{1+\frac{2}{3}\omega }-1\right) ^{p_{-}}\nonumber\\
&\times&\left( 9+6\omega \right) ^{-%
\frac{1}{4+3\omega }}t_{m}^{-\frac{2}{4+3\omega }}\;.  
\end{eqnarray}
Now, it is possible to have a transition from a GR behaviour to a BD behaviour during a single phase of the evolution of the universe \cite{chauvineau}. The success of the GR theory concerning the structure formation can be kept. But, new features (accelerated expansion?) may be introduced non-trivially. Work is in progress in order to implement explicitly possible realistic scenarios using this structure.

\section{Quantum cosmology}

A simplified approach to a quantum cosmological model of the Brans-Dicke theory can be made by exploring the Brans-Dicke action in the mini-superspace. In such approach, the cosmological metric is used, and the only dynamical variables are the scale factor and the scalar field. The metric is written as
\begin{eqnarray}
ds^2 = N^2dt^2 - a^2(t)(dx^2 + dy^2 + dz^2)\;,
\end{eqnarray}
where $N$ is the lapse function which reflects the invariance of the theory with respect to time-reparametrisation in the geometric theories of gravity. The lapse function determines the Hamiltonian constraint.

With these considerations, the gravitational action of the Brans-Dicke theory (\ref{bd}) reads:
\begin{equation}
\mathcal{L}_{G}=\frac{1}{N}\left\{ 6\left[ \varphi a\dot{a}^{2}+a^{2}\dot{a}%
\dot{\varphi}\right] -\omega a^{3}\frac{\dot{\varphi}^{2}}{\varphi }\right\}\;,
\label{Lagrangian 2}
\end{equation}%
where we have already discarded the surface terms. The Lagrangian can be written as function of the conjugated momenta, defined by:
\begin{eqnarray}
\pi_q = \frac{\partial \mathcal{L}}{\partial \dot q}\;.
\end{eqnarray} 
Using this expression, we obtain the following Hamiltonian:
\begin{equation}
\mathcal{H}=N\left\{ \frac{1}{\left( 3+2\omega \right) }\left[ \frac{\omega 
}{12\varphi a}\pi _{a}^{2}+\frac{1}{2a^{2}}\pi _{a}\pi _{\varphi }-\frac{%
\varphi }{2a^{3}}\pi _{\varphi }^{2}\right] -\frac{1}{a^{3\alpha }}\pi
_{T}\right\}\;,  
\label{Total Hamillt.}
\end{equation}%
Remark that effectively the lapse function implies the Hamiltonian constraint.
In this equation, $\pi_T$, $\pi _{a}$ and $\pi _{\varphi }$ are the conjugate
momenta associated with the matter component, the scale factor $a$ and the field $\varphi$, respectively. We have introduced the matter component, defined by the equation of state $p = \alpha\rho$, using Schutz's formalism (for details on how to compute the expression for the conjugate momentua, see Refs. \cite{demaret, rubakov, lemos1, lemos2}). In fact, using the approach employed in those references, the matter Hamiltonian becomes: 
\begin{equation}
\mathcal{H}_{M}=-\frac{N}{a^{3\alpha }}\pi _{T}\;,  
\label{Hamilt. matter}
\end{equation}%
The quantisation of the equation (\ref{Total Hamillt.}), together with the
Wheeler-DeWitt equation, results in a Schr\"odinger-like equation, from where
we obtain a Hamiltonian operator. The canonical quantisation $\pi _{k}\rightarrow -i\partial _{k}$, including the ambiguity due to ordering factors $\bar{p}$ and $\bar{q}$, gives us:
\begin{equation}
\frac{\omega }{12\varphi }\left( \frac{1}{a}\partial _{a}^{2}\Psi +\frac{%
\bar{p}}{a^{2}}\partial _{a}\Psi \right) +\frac{1}{2a^{2}}\partial
_{a}\partial _{\varphi }\Psi -\frac{1}{2a^{3}}\left( \varphi \partial
_{\varphi }^{2}\Psi +\bar{q}\partial _{\varphi }\Psi \right) =\frac{i\left(
3+2\omega \right) }{a^{3\alpha }}\partial _{T}\Psi\;.
\label{Schroed. 1}
\end{equation}%
In order to avoid the cross derivative $\partial _{a}\partial _{\varphi }$,
let us make a change of coordinates:
\begin{equation}
a\rightarrow \varphi ^{-\frac{1}{2}}b;\qquad \varphi \rightarrow \varphi\;.
\label{coordinates transf.}
\end{equation}
With those new coordinates and redefining $T\rightarrow T/24$, equation (\ref{Schroed. 1}) becomes:
\begin{equation}
-\left\{ \frac{\varphi ^{\frac{1}{2}}}{b}\left[ \partial _{b}^{2}+\frac{p}{b}%
\partial _{b}\right] -\frac{12\varphi ^{\frac{3}{2}}}{b^{3}}\left[ \varphi
\partial _{\varphi }^{2}+q\partial _{\varphi }\right] \right\} \Psi =\frac{%
\varphi ^{\frac{3\alpha }{2}}}{b^{3\alpha }}i\partial _{T}\Psi\;,
\label{Schroed. 2}
\end{equation}
with new ordering factors, which are related to the previews ones in the following way:
\begin{equation}
p=\frac{2\omega \bar{p}+9-6\bar{q}}{3+2\omega };\qquad q=\bar{q}\;.
\label{p,q relations}
\end{equation}
The coordinate transformations (\ref{coordinates transf.}) correspond in fact to a conformal transformation that implies to rewrite the Brans-Dicke action in the minimal-coupling frame. In order to keep a simple framework, the matter sector will be that of a radiative fluid, which is conformally invariant.

One can verify that the Hamiltonian operator
\begin{equation}
\hat{H}=-\frac{b^{3\alpha }}{\varphi ^{\frac{3\alpha }{2}}}\left\{ \frac{%
\varphi ^{\frac{1}{2}}}{b}\left[ \partial _{b}^{2}+\frac{p}{b}\partial _{b}%
\right] - \frac{12}{(3+2\omega)} \frac{\varphi ^{\frac{3}{2}}}{b^{3}}\left[ \varphi \partial
_{\varphi }^{2}+q\partial _{\varphi }\right] \right\}\;,
\label{Hamiltonian operator}
\end{equation}%
coming from equation (\ref{Schroed. 2}) is symmetric (or Hermitian, an equivalent terminology), such that:
\begin{equation}
\left\langle \phi ,H\psi \right\rangle = \left\langle H\phi ,\psi \right\rangle\;,
\end{equation}
in its domain only if we consider the measure
\begin{equation}
\left\langle \phi ,\psi \right\rangle =\int_{0}^{\infty }\int_{0}^{\infty
}\left( \overline{\phi }\psi \right) b^{p-3\alpha +1}\varphi ^{q+\frac{%
\left( 3\alpha -5\right) }{2}}dbd\varphi\;.  
\label{measure}
\end{equation}
Notice that this measure in the Hilbert space depends on the ordering factors. 

Even if the Hamiltonian is symmetric, this fact does not guarantee that it is self-adjoint. To establish the self-adjoint properties of the Hamiltonian (\ref{Hamiltonian operator}), it is necessary to determine the von Neumann deficiency indices. This may depends on the factoring order, but in most of the cases it happens that the Hamiltonian (\ref{Hamiltonian operator}) is in fact self-adjoint, and in the cases where it is not already self-adjoint, it admits self-adjoint extension \cite{carla}.

One important feature connected with the Hamiltonian (\ref{Hamiltonian operator}): its energy spectrum is positive defined only if $\omega < - \frac{3}{2}$. Hence, consistent quantum models seem to be possible only for the phantom regime. In this case, it is possible to obtain a prediction for the evolution of the universe. In Ref. \cite{moniz} the expectation value for the scale factor has been evaluated, displaying a singularity-free universe, with a bounce in the scale factor.

\section{Conclusions}

The Brans-Dicke theory has appeared in the sixties as an important alternative to the GR theory. Later, the interested in this proposal has decreased considerably due to the restriction imposed by the local tests represented, for example, by the PPN analysis, which reduced the BD theory essentially to the GR theory. However, the connections of the BD theory to multidimensional and string theories have kept the BD theory a strongly investigated proposal. Recently, its connection with the $f(R)$ theories, which is equivalent to the BD theory with a potential, gave a revival of this prototype of scalar-tensor theory. Again, problems with local tests appeared which could be cope with through some mechanisms, like the chameleon one.

The frequent mention to the BD theory in the literature shows that it contains yet many unexplored features that could lead to a solid alternative to the GR theory. We must remember that in spite of its many success, particularly for small scale gravitational phenomena, the GR theory faces huge problems in explaining large scales phenomena, besides the difficulties of the appearance of singularities in many different, but important, contexts.

In the present text, we tried to show the richness of some unexplored feature of the BD theory, like its black hole type structures when the energy conditions are violated, and the general cosmological solutions, at least in the flat case. Moreover, we have described a new view of the BD theory based on considering the scalar field as an external scalar field. This proposal opens news possibilities to construct a viable cosmological scenario, and new static, spherically symmetric configurations. Work is in progress in order to determine specific viable models.

Some features of quantisation of the Brans-Dicke action in the mini-superspace, with the time variable being recovered by employing tSchutz's description for a perfect fluid: the phantom case, represented by $\omega < - \frac{3}{2}$ seems to lead to a consistent quantum version, while the ordinary scalar field case leads to many anomalies, specially the absence of positive defined energy; moreover, a non trivial measure must be introduced in order to construct a self-adjoint operator in both cases. 

The richness of this scalar-tensor theory, already more the half century old, seems not have been exhausted yet.

\bigskip

\noindent
{\bf Acknowledgements}: We thank CNPq (Brazil) and FAPES (Brazil) for partial financial support.

\end{document}